# Comparative Analysis of Different Techniques of Real Time Scheduling for Multi-Core Platform

Girish Talmale* and Urmila Shrawankar**
*Assistant Professor,GHRCE Nagpur
girishtalmale@gmail.com
**Associate Professor,GHRCE,Nagpur
urmilla@gmail.com

**Abstract:** As the demand of real time computing increases day by day, there is a major paradigm shift in processing platform of real time system from single core to multi-core platform which provides advantages like higher throughput, linear power consumption, efficient utilization of processor cores and high performance per unit cost over the many single core processors unit. Currently available most popular real time schedulers for multi-core domain are partitioned and global scheduling and these schedulers not suitable to efficiently use this multi-core platform efficiently. Although, semi-partitioned algorithms increases utilization bound by using spare capacities left by partitioning via global scheduling, it has a inherent disadvantage of off-line task splitting. Although, semi-partitioned algorithms increases utilization bound by using spare capacities left by partitioning via global scheduling, it has a inherent disadvantage of off-line task splitting. To overcome these problems of multi-core real time scheduling algorithm new dynamic cluster based multi-core real time scheduling algorithm proposed which is hybrid scheduling approach. This paper discuss different multi-core scheduling techniques and comparative analysis of these techniques with the proposed dynamic cluster based real time multi-core scheduling.

## 1.Introduction

Real-time systems have been pervasive in our daily life. Recent advancements of multi-core architectures have led to increasing use of multi-cores in real-time systems rather than building more complicated single core architecture and raising its working frequency[1]. Multi-core is becoming mainstream and provides advantages like higher throughput, linear power consumption, efficient utilization of processor cores and high performance per unit cost over the many single core processors unit over single core architecture as it suffer from high clock frequency ,more heat dessipation ,small memory size and less memory access speed.The miltiprocessor performance depends upon the type of application and the implementation of software.To exploit the concurrencay offerd by the multicore architecture ,suitable algorithm is use that divedes the application software into tasks and efficient scheduling techniques is required that fairly distributes tasks to processors.Recent advancement in multicore platform and rapidly incresing demand of real time computation led to regain interest of an multicore real time scheduling.
This paper present comparative analysis of different multi-core scheduling algorithm and resource sharing protocols. The organization of this paper as follows section 2 provides the different multi-core real time scheduling algorithms .Section 3 describe the comparative analysis of different multi-core scheduling algorithm and section 4 describe conclusions of study

## 2.Multi-core Scheduling Algorithms

The multi-core scheduling is basically used to solve the following problems. The problem associate with  how to assign task to processor. The second problem is associate with priority that decides when and in what order the jobs of the tasks executes on the processor. The multi-core scheduling algorithms are classified on the basis of migration and priority of tasks. Scheduling algorithm where no migration is allowed is refer as partitioned and where the migration is allowed is termed as global. The hybrid scheduling algorithms use both the approaches where some task are not allowed to migrate and some like portioned scheduling and some task are allowed to migrate like global scheduling approach. The hybrid multi-core scheduling are semi partitioned and cluster based scheduling. The cluster based scheduling approach further enhance by using static and dynamic cluster based scheduling.

## 2.1 Partitioned Scheduling

In partitioned scheduling each task is statically assign to dedicated processor and no migration of task is allowed as shown in Figure1 .Every processor having their own scheduler with separate task run queue and no migration is allowed during run time. The main benefits of using partitioned scheduling is that ones the task allocate to a particular processor ,the existing uniprocessor scheduling algorithms can be applied for scheduling the task to the processor. The second advantages for using partitioned scheduling is no migration overhead ,as the task are not allowed to migrate during run time like global scheduling approach. The another benefits of partitioned approach is simple queue management as each processor has its own ready queue. The system utilization is low under partitioned scheduling and if the total utilization of the task set reaches slightly higher than 50% then then the deadline may be missed.

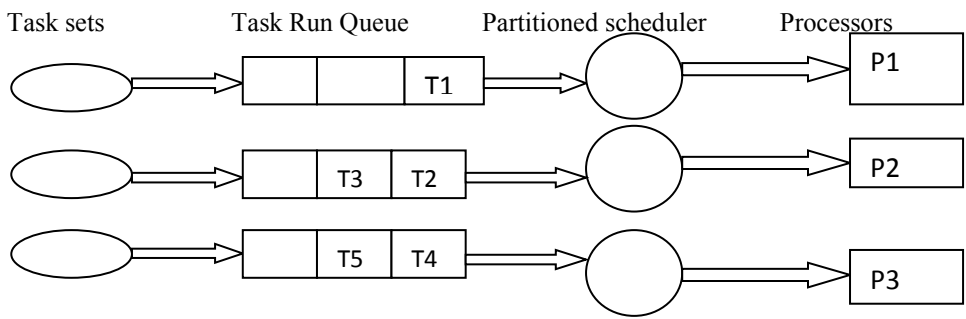

Figure1. Partitioned Scheduling

In the above figure the task T1 is statically assigned to processor P1 ,task set containing tasks T2 and T3 are statically assigned to processor P2 and task T4 and T5 are statically assigned to processor P3 using petitioned scheduling. However to assigned these task sets to processor so that no processor get overloaded is major problem in partitioned scheduling and that task allocation problem is analogous to bin packaging problem which is known as NP hard problem. The portioned algorithm suffers from low system utilization and poor load balancing as the partitioned scheduling algorithm cannot guarantee to portioned a task set with total utilization greater than (M+1)/2 on platform consist of total M processors because the processor may be idle but cannot be allocated to ready task that are statically assigned to different processors .Partitioned scheduler require more number of processor to schedule the task sets when compare to global scheduler.

The partitioned scheduling algorithm initially used existing uniprocessor scheduling algorithm such as P-EDF(Partitioned-Earliest Deadline First)[2][3][4] as dynamic priority based partitioned scheduling algorithm and P-RM(Partitioned-Rate Monotonic)[7] as fixed priority based partitioned scheduling algorithm with bin packing heuristics such as First Fit,Next Fit ,Best Fit and Worst Fit and task ordering such as decrease utilization for task allocation. To improve performance of dynamic and fixed priority based partitioned scheduling algorithm, different variants of EDF such as EDF-US(EDF-utilization separation) [8] and EDF-FFID(EDF-First Fit Increasing Deadline)[23] scheduling algorithms are proposed.

## 2.2 Global Scheduling

In global scheduling task are globally assigned to all processors. All task are maintained in one priority queue and may migrate among all processors as shown in fig.2 In the following figure the task T1,T2,T3 and T4 are maintained in the global priority queue and assigned all processor P1,P2 and P3. These task can migrate from one processor to another.

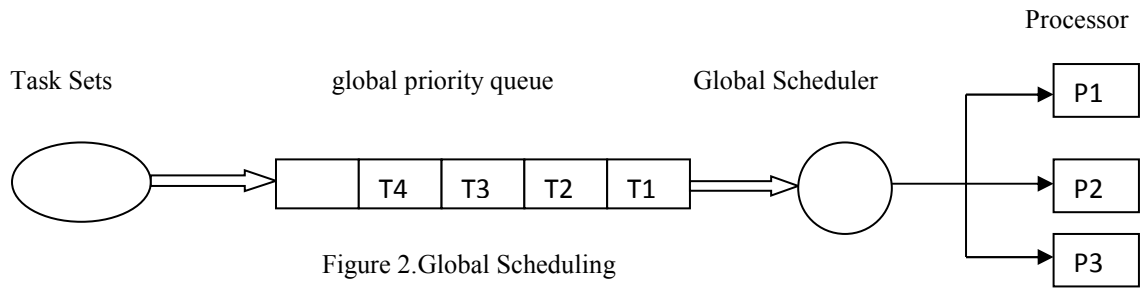

Figure 2.Global Scheduling

The main advantages of global scheduling is that it solves the problem of task assignment in partitioning scheduling as all the task are assigned to all processors in the system .The second advantages of global scheduling is less scheduling overhead due to preemption of task as the task are preempted only when there is no idle processor. Global scheduling is more suitable for open system where the new task can arrive dynamically and assigned to the existing schedule without assigned it to a particular processor like partitioned scheduling.

The example of global scheduling is G-EDF[1][2][21][23](Global-Earliest Deadline First) .There are many variants of global scheduling algorithm  such as G-LLF(Global Least Laxity First) EDZL(Earliest Deadline until Zero Laxity)[17][8] EDCL(Earliest Deadline until critical laxity) LLREF(Largest Local Remaining Execution Time First)[20]based on laxity or remaining execution time etc, which improve schedulability of the task sets under global scheduling .These global scheduling algorithm are not optimal. Proportionate Fair (P-fair) global scheduling algorithm ensures proportionate fairness means that every active job get equal fair of processor time. The another family of P-fair algorithm based on cache aware policy was proposed that allows scheduling of independent task sets on multi-core platform. Boundary Fair global scheduling algorithm reduce scheduling points of P-Fair by making scheduling decision only at the period boundaries or deadlines .All these P-fair algorithms are based on discrete time model that's why task never execute for less than one system time period. In continuous time model, task execution not synchronized on system time unit therefore task can execute for any period of time. DP-Fair (Deadline Partitioned Fair) algorithm was proposed which is boundary fair algorithm based on continuous time model. All the above mentioned algorithm suffer from some common drawbacks such as uniprocessor scheduling cannot directly use in global scheduling like partitioned scheduling. Global scheduling increases job migration overhead as the jobs can migrate among processors. The another problem in global scheduling is the management of global data structure for the task queue as all the task are maintained in global priority queue.

## 2.3 Semi-Partitioned Scheduling

In semi-partitioned scheduling is a type of hybrid scheduling approach in which some task are assigned to a specific processor like partitioned scheduling and some task are globally assigned to set of processor like global scheduling as shown in following figure.

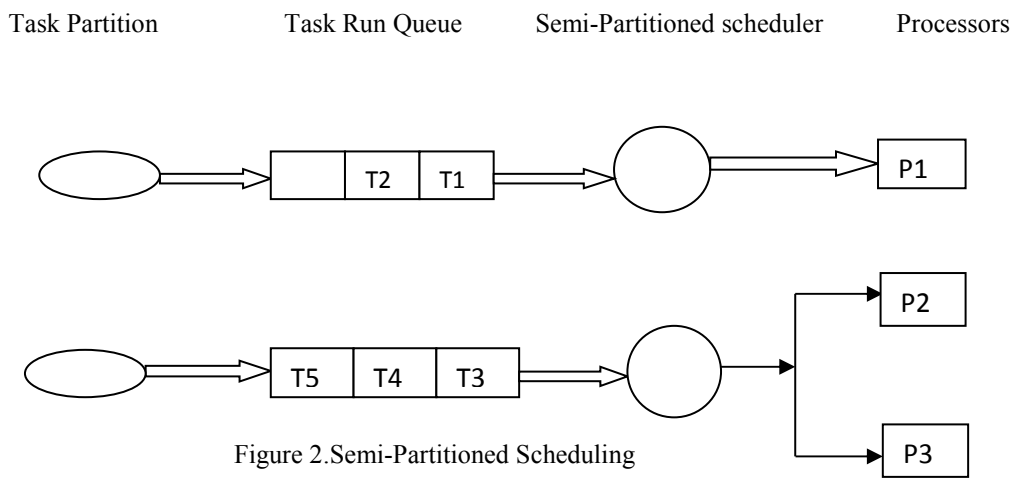

Figure 2.Semi-Partitioned Scheduling

In the above fig.of semi-partitioned scheduling task T1 and T2 are scheduled to statically to fix processor P1 like partitioned scheduling  and tasks T3,T4 and T5 are globally scheduled to processor P2 and P3 like global scheduling. In semi-partitioned scheduling most of the tasks are executed on single processor and few no of task are allowed to migrate among the set of processor.The advantages of this scheduling techniques is to improve processor utilization of partitioned scheduling by globally scheduling the tasks that cannot be assigned to one processor due to bin packaging heuristics. The tasks that cannot be completely assigned to one processor will be split up and allocate to different processors. This processes of assigned the task to processor is offline. The example of semi-partitioned scheduling is EKG[2][9](EDF with task splitting and k processor in a group).This is the optimal semi-partitioned scheduling algorithm for periodic task sets under implicit deadline. The tasks that are globally assigned are refer as the migratory tasks and the tasks which are assigned to single processor are refer as component tasks. Another semi-petitioned algorithm is EDDP(Earliest Deadline Deferrable Portion)[10][13] [14] places each task with utilization greater than 65% on its own processor and the tasks with low utilization assigned to set of processor with tasks are divided into two portion and these two portion of split task are not allowed to execute simultaneously by differing the execution of the portion of the tasks on lower number of processor while the portion on higher number of processor executes. Partitioned Deadline Monotonic Scheduling Highest Priority Task Split(PDMS HPTS)[21] is the semi-partitioned scheduling algorithm based on fixed priority scheduling with sporadic task sets with implicit or constrained deadline. It split only single tasks on each processor. The problem in semi-partitioned scheduling is offline task splitting. The drawbacks of semi-partitioned scheduling is poor processor utilization and load balancing as compare to global scheduling.

## 2.4 Cluster Scheduling

Cluster scheduling is combination of partition and global scheduling where task are divided into sets and these set of tasks are allocated to cluster which consist of set of processors and are scheduled globally within the cluster as shown in fig 4.

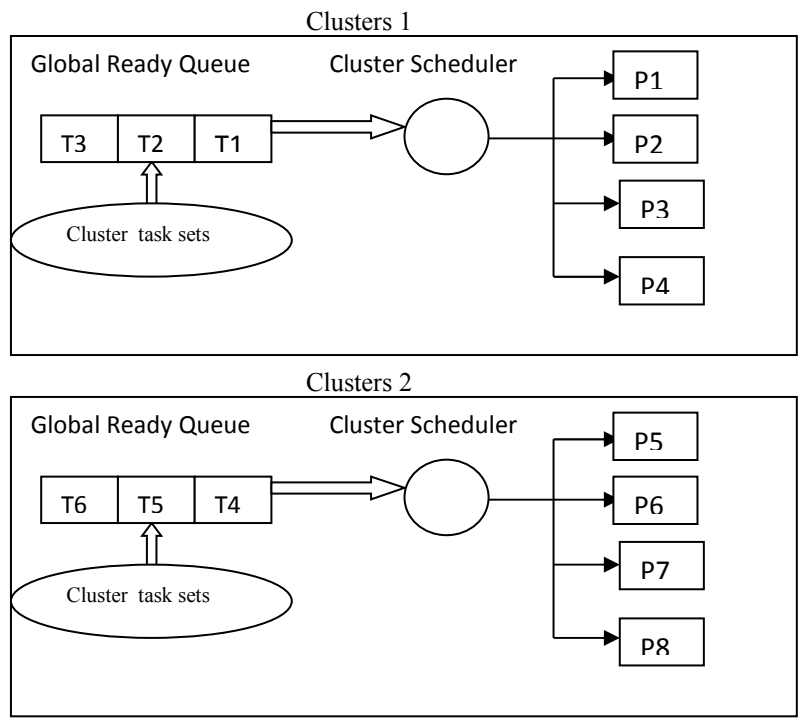

Figure.4 Cluster Scheduling

In cluster scheduling the set of m processors are divided into ( m/c) set of c processor each .If c=1 then cluster scheduling becomes partitioned scheduling and if c=m then cluster scheduling becomes global scheduling. It simplify the bin packing problem of partitioned scheduling by distributing the task to clusters. Cluster scheduling can improve the utilization; reduce overhead due to migration and response time by applying different heuristics on cluster scheduling. Different type of cluster can be form based on high and low utilization of tasks. Another flexibility offered by the clustering is to create cluster with different resource capability. Clustered scheduling is likely to grow in importance as multi-core platforms become larger and less uniform. Cluster based scheduling can be classified into two types static and dynamic clustering .Under static clustering, each cluster is assigned to a fix set of processor like one to one mapping as shown in fig.5

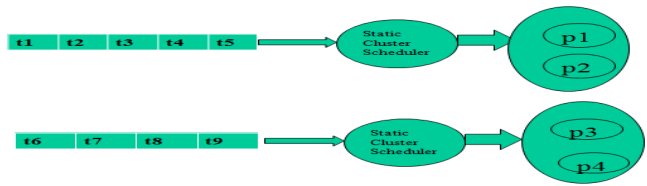

Figure.5 Static Cluster Scheduling

In dynamic cluster scheduling cluster are assigned to processor dynamically like one to many mapping as shown in fig.6

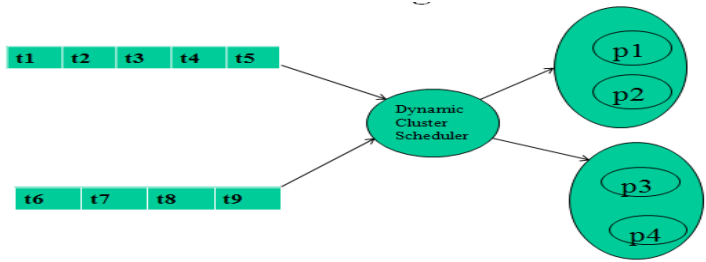

Figure 6. Dynamic Cluster Scheduling

Dynamic cluster scheduling is more general and less sensitive to task cluster mapping compare to static clustering

## 3. Comparative Analysis of different scheduling algorithms

| Multi-core Scheduling | Utilization Bound | Approximation Ratio | Resource augmentation | Average Response Time | Load Balancing | Resource Reliability | Open System Compatibility | Multi-core with large no of cores adaptability | Scheduling Overhead | Job Migration Overhead | Implementation Complexity |
|---|---|---|---|---|---|---|---|---|---|---|---|
| Partitioned Scheduling | Very Low | Very high | Very low | Low | Poor | Very Low | Very Low | Low | Very Low | Very Low | Very Low |
| Global Scheduling | Very High | High | Low | High | Best | Low | Very High | Very Low | High | Very High | Very High |
| Semi-Partitioned Scheduling | Average | Average | Average | Average | Average | Average | Average | Average | Average | High | High |
| Static Cluster Scheduling | High | Low | High | High | Good | Good | High | High | Average | Average | Very High |
| Dynamic Cluster Scheduling | Very high | Very Low | Very High | Very High | Best | Very High | Very High | Very High | High | High | Very High |

## Conclusion

Multi-core processor now days extensively used in real time system to satisfy high real time computational demand. To exploit potential of multi-core platform, efficient multi-core real time scheduling is required. The two most popular techniques for multi-core scheduling are partitioned and global scheduling. Partitioned scheduling suffer from waste of resource capacity, task splitting problem while global scheduling suffer from high overhead for migrating task, management of large queue data structure. The semi-partitioned based scheduling approach is used to overcome this problem but it suffers from low resource utilization and offline task splitting. The proposed cluster based scheduling is a hybrid approach that combining benefits of both partitioned and global scheduling. The cluster based scheduling further classified into static and dynamic cluster scheduling. In static cluster based scheduling each cluster is assigned to a fix set of processor. In dynamic cluster based scheduling cluster dynamically assigned to processor. Comparitive analysis of these scheduling algorithms shows that dynamic cluster scheduling offer several advantages over existing scheduling approaches in terms of resource utilization, load balancing, reliability etc.